\documentclass[a4paper,fleqn]{cas-sc}

\usepackage[numbers]{natbib}
\usepackage{xcolor}
\usepackage{verbatim}
\usepackage{tcolorbox}
\usepackage{hyperref}
\usepackage{courier}
\usepackage{caption}
\usepackage{subcaption}
\usepackage{amsmath}
\usepackage{booktabs}
\usepackage{pifont}
\usepackage{float} 
\usepackage{algorithm}
\usepackage{algorithmic}

\usepackage{tabularx,colortbl}
\usepackage{multirow}
\usepackage{times}
\usepackage{array}
\usepackage{rotating}

\usepackage{graphicx}
\usepackage[flushleft]{threeparttable}
\usepackage{todonotes}
\usepackage[export]{adjustbox}[2011/08/13]
\usepackage{longtable}

\hypersetup{
colorlinks=true,
linkcolor=blue,
anchorcolor=blue,
urlcolor=blue,
citecolor=blue,
}

\definecolor{softgreen}{rgb}{0.0, 0.5, 0.3}

\def\tsc#1{\csdef{#1}{\textsc{\lowercase{#1}}\xspace}}
\tsc{WGM}
\tsc{QE}
\tsc{EP}
\tsc{PMS}
\tsc{BEC}
\tsc{DE}

\begin{document}
\let\WriteBookmarks\relax
\def\floatpagepagefraction{1}
\def\textpagefraction{.001}
\shorttitle{Data Preparation for Deep Learning based Code Smell Detection: A Systematic Literature Review}
\shortauthors{Fengji Zhang et al.}

\title [mode = title]{Data Preparation for Deep Learning based Code Smell Detection: \\ A Systematic Literature Review}

\author{Fengji Zhang\textsuperscript{\textit{a }}}
\ead{fengji.zhang@my.cityu.edu.hk}
\address{\textsuperscript{\textit{a}} Department of Computer Science, City University of Hong Kong, Hong Kong, China}

\author{Zexian Zhang\textsuperscript{\textit{b,c}}}
\ead{zexianzhang@whut.edu.cn}
\address{\textsuperscript{\textit{b}} School of Computer Science and Artificial Intelligence, Wuhan University of Technology, Wuhan, China}

\address{\textsuperscript{\textit{c}} Sanya Science and Education Innovation Park of Wuhan University of Technology, Sanya, China}

\author{Jacky Wai Keung\textsuperscript{\textit{a}}}
\ead{jacky.keung@cityu.edu.hk}

\author{Xiangru Tang\textsuperscript{\textit{d}}}
\ead{xiangru.tang@yale.edu}
\address{\textsuperscript{\textit{d}} School of Engineering \& Applied Science, Yale University, New Haven, United States}

\author{Zhen Yang\textsuperscript{\textit{e}}}
\ead{zhenyang@sdu.edu.cn}
\address{\textsuperscript{\textit{e}} School of Computer Science and Technology, Shandong University, Tsingtao, China}

\author{Xiao Yu\textsuperscript{\textit{b,f}}}
\ead{xiaoyu@whut.edu.cn}
\address{\textsuperscript{\textit{f}} Wuhan University of Technology Chongqing Research Institute, Chongqing, China} 

\author{Wenhua Hu\textsuperscript{\textit{b}}}
\ead{whu10@whut.edu.cn}

\begin{abstract}
\underline{C}ode \underline{S}mell \underline{D}etection (CSD) plays a crucial role in improving software quality and maintainability. And \underline{D}eep \underline{L}earning (DL) techniques have emerged as a promising approach for CSD due to their superior performance. However, the effectiveness of DL-based CSD methods heavily relies on the quality of the training data. Despite its importance, little attention has been paid to analyzing the data preparation process.
This systematic literature review analyzes the data preparation techniques used in DL-based CSD methods. We identify 36 relevant papers published by December 2023 and provide a thorough analysis of the critical considerations in constructing CSD datasets, including data requirements, collection, labeling, and cleaning. We also summarize seven primary challenges and corresponding solutions in the literature.
Finally, we offer actionable recommendations for preparing and accessing high-quality CSD data, emphasizing the importance of data diversity, standardization, and accessibility. This survey provides valuable insights for researchers and practitioners to harness the full potential of DL techniques in CSD.
\end{abstract}

\begin{keywords}
Code Smell Detection,
Deep Learning,
Data Preparation,
Systematic Literature Review
\end{keywords}

\maketitle
\section{Introduction} \label{Introduction}
Code smell refers to certain symptoms or indications in the source code that suggest there may be underlying problems or potential design flaws
~\citep{danphitsanuphan2012code,santos2018systematic,zakeri2023systematic, li2023relative}. 
It does not necessarily indicate a functional error or bug in the code but rather highlights programming practices that can impair the maintainability, readability, and extensibility of the software 
~\citep{di2018detecting,alazba2023deep, hu2023revisiting, liu2024revisiting}.
\underline{C}ode \underline{S}mell \underline{D}etection (CSD) aims to automatically identify code smells in software source code to ensure code quality, improve software maintainability, and promote good programming practices. 
Recently, \underline{D}eep \underline{L}earning (DL) techniques are gaining popularity in the CSD task~\citep{guo2019deep,kim2020deep,hamdy2020deep}. The main advantage of DL models is their ability to automatically encode and learn from raw data, eliminating the need for handcrafted rules and feature engineering presented in previous heuristic-based and machine learning-based CSD methods \citep{sharma2018survey,alkharabsheh2019software,jain2021improving}. Despite the outstanding performance of the DL-based CSD methods, they require a substantial amount of training data to model the complexity of code smells~\citep{di2018detecting}. 
High-quality training data plays a key role in the validity of model results. Noisy datasets can hinder effective model training and affect result reliability~\citep{fakhoury2018keep,ardimento2021temporal}. Data quality could also impact the scalability of models~\citep{allal2023santacoder}. Early decisions when constructing CSD datasets, such as the choice of language~\citep{virmajoki2022detecting,siddiq2022empirical} and application scenarios~\citep{zhang2022delesmell,kaur2023improving}, can affect the scaling and generalization ability of models. Consequently, the availability of high-quality code smell datasets is crucial for building effective DL-based CSD models.

Despite the critical importance of data to CSD models, little attention has been paid to systematically analyzing the CSD data preparation process. Recent literature surveys \citep{alazba2023deep, naik2023deep, malhotra2023examining} comprehensively analyzed DL-based CSD methods. 
They covered many facets, including code smell types, deep learning techniques, model features, evaluation methods, and the datasets used. They indicated insights on programming language preferences, model effectiveness, and dataset characteristics. 
However, they overlooked crucial aspects of data preparation, such as requirements, collection, cleaning, and labeling techniques. 
Consequently, they lack a comprehensive view of data preparation challenges and strategies, limiting researchers and practitioners in harnessing the complete potential of DL techniques in CSD.

To address these gaps, this survey systematically analyzes the existing data preparation processes for DL-based CSD. We identify relevant papers through a \underline{S}ystematic \underline{L}iterature \underline{R}eview (SLR) process, collecting 36 papers on DL-based CSD studies published until December 2023. We then carefully analyze the collected papers concerning data preparation considerations, encountered challenges, and proposed solutions. Additionally, we provide recommendations for preparing and accessing high-quality CSD data.
This survey is organized around three main \underline{R}esearch \underline{Q}uestions (RQs):

\begin{itemize}
    \setlength{\itemsep}{0pt}
    \setlength{\parsep}{0pt}
    \setlength{\parskip}{0pt}
    \item[RQ1] \textit{What are the critical considerations in constructing CSD datasets?} This research question aims to understand the critical factors researchers consider when building DL-based code smell detection datasets. We analyze papers regarding the four main phases of the established machine learning workflow~\citep{amershi2019software}: data requirements, collection, labeling, and cleaning. For data requirements, we examine the programming language, code smell types, and detection scenarios addressed. For data collection, we analyze the data sources and types. For data labeling, we summarize the costs and efficiency of automatic, manual, and semi-automatic approaches. Finally, for data cleaning, we identify issues of code noise and redundancy.

    \item[RQ2] \textit{What are the challenges in existing CSD datasets?} This research question identifies challenges that may hinder the performance and reliability of DL-based CSD methods due to data issues. We analyze seven primary challenges: data scarcity, limited generalization, inaccessibility, heavy expert dependency, difficulty in labeling, data imbalance, and redundancy.

    \item[RQ3] \textit{What are the solutions presented in the literature?} Given the challenges identified in RQ2, this research question summarizes solutions proposed in the literature. To the challenges addressed, we map five approaches - cross-project datasets, two-phase data utilization, resampling, semi-automatic labeling, and data cleaning methods.
    
\end{itemize}

Finally, we provide recommendations based on this survey. Future work should focus on creating more diverse, publicly available datasets that address current limitations. Researchers could leverage multiple programming languages, data sources, and domains to improve generalizability. Semi-automatic labeling and automated real-world data collection may help scale datasets while maintaining quality. Adopting best practices for data governance, including documenting the data collection and pre-processing details, would enhance transparency and reproducibility. Establishing standard criteria to evaluate datasets could help standardize their construction and quality assessment. These efforts aim to generate larger, higher-quality datasets allowing DL-based models to better learn complex code smell patterns across different application scenarios.

The main contributions of this research are:
\begin{itemize}
\setlength{\itemsep}{0pt}
\setlength{\parsep}{0pt}
\setlength{\parskip}{0pt}
\item Introduce the first systematic review of data preparation processes for DL-based CSD methods  (RQ1 in Section~4).
\item Provide thorough solutions mapped to identified data challenges to guide future dataset preparation (RQ2 in Section~5 and RQ3 in Section~6).
\item Propose recommendations on diversifying and standardizing datasets through multi-language modeling, semi-automatic labeling, and best practices for data governance and accessibility (Section~7).
\end{itemize}

The rest of the paper is organized as follows. Section \ref{Background and Related Work} describes background and related work. Section \ref{SLR} details our SLR methodology. Section \ref{RQ1} to \ref{RQ3} presents results addressing each RQ. Section \ref{recommendation} provides recommendations based on findings. Section \ref{threat} discusses threats to validity.  Section \ref{conclusion} concludes the paper.

\section{Related Work} \label{Background and Related Work}
\begin{figure}[t]
 \centering
 \includegraphics[width=0.5\linewidth]{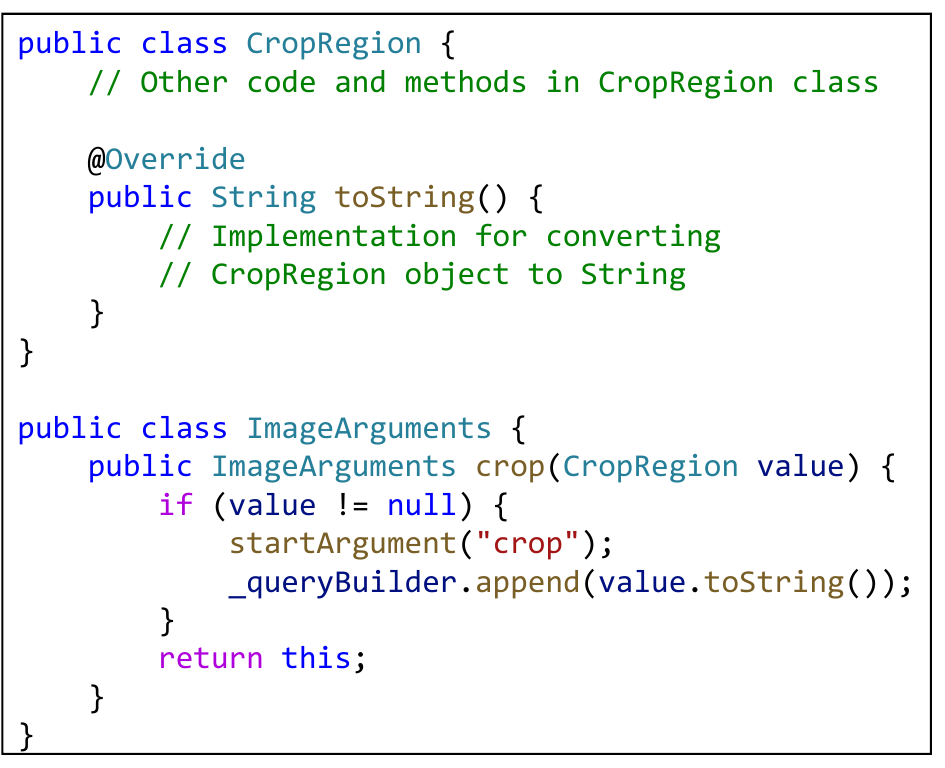}
 \caption{An example of \textit{Feature Envy} code smell.}
 \label{Example}
\end{figure}

This section provides context on code smells, code smell detection techniques, and related CSD SLRs. The background informs the goals and context of our study.

\subsection{Code Smell Detection}
Code smells represent undesirable design or implementation constructs that can degrade code maintainability and quality, indicating structural patterns correlated with increased defect risk~\citep{al2020bad, fowler2018refactoring, kim2017finding}. One example is the \textit{Feature Envy} shown in Figure \ref{Example}. In this case, the \textit{crop} method belongs to the \textit{ImageArguments} class but is coupled to the \textit{CropRegion} class. The \textit{CropRegion} class has a \textit{toString()} method that converts the \textit{CropRegion} object to a string. While the \textit{crop} method calls \textit{value.toString()}, which focuses too much on the internal details of the \textit{CropRegion} class about its string conversion logic. Since \textit{crop} is defined in \textit{ImageArguments}, it should not read and manipulate attributes of \textit{CropRegion} rather than its owning class. The \textit{crop} method would be better defined as a method in \textit{CropRegion} since it primarily operates on that class's data rather than \textit{ImageArguments}. This excessive dependency on another class's implementation indicates the presence of \textit{Feature Envy}, which can negatively impact code understandability, modification, and testing.

Early code smell detection approaches are generally heuristic-based or machine learning-based models \citep{sharma2018survey,alkharabsheh2019software}. 
However, heuristic-based methods are criticized for their subjectivity due to manually adjusted heuristic rules~\citep{di2018detecting}. Machine learning-based methods require careful construction and selection of code smell features~\citep{fontana2017code}, which also heavily relies on human expertise.

With advancements in deep learning technology in the field of both artificial intelligence and software engineering  ~\cite{chen2023deep,yang2023significance,gao2023rumor,chen2020deep,ma2023attsum,qiao2023effective}, recent researchers have introduced various deep learning techniques for automatically extracting code smell features from code \citep{tarwani2022application}.  Model architectures like convolutional neural networks  \citep{fakhoury2018keep,das2019detecting,hamdy2020deep,yin2021local}, long short-term memory networks \citep{guo2019deep,wang2020feature,yu2021novel,ardimento2021temporal,li2022multi,siddiq2022empirical,ho2023fusion}, and recurrent neural networks \citep{das2019detecting,hamdy2020deep,siddiq2022empirical} have achieved state-of-the-art CSD accuracy \citep{zhang2022delesmell}.

\subsection{Related CSD SLRs}

Despite data preparation playing a pivotal role in CSD, comprehensive investigations have been lacking. Several surveys in CSD methods explored machine learning and deep learning techniques. While these surveys discussed CSD datasets to varying extents, most of them lacked detailed examination of the data preparation process and systematic identification of challenges or solutions.   
Specifically, \citet{gupta2017systematic} delivered a comprehensive review of CSD studies from 1999 to 2016, stressing the pivotal role of code smells in software maintenance. 
\citet{azeem2019machine} and \citet{al2020bad} explored machine learning-based CSD studies until 2018, focusing on the types and performance of machine learning techniques. They both found that the random forest emerged as the most effective technique for detecting various code smells and emphasized the significance of manually validating datasets, noting a scarcity of available datasets.
\citet{kaur2021review} reviewed CSD studies up to 2020, concentrating on simple and hybrid machine learning techniques and their evaluation methods. They revealed that support vector machine and decision tree algorithms were frequently used by the researchers, and much of the research focused on open-source software. Additionally, they noted that most of the researchers used small and medium-sized datasets and lacked valid industrial datasets.
\citet{lewowski2022far} assessed the reproducibility of CSD research from 1999 to 2020, focusing on machine learning-based studies. 

\citet{alazba2023deep}, \citet{naik2023deep}, and \citet{malhotra2023examining} are all devoted to systematically reviewing the research progress in the field of DL-based CSD. They provided a comprehensive survey and summary of the developments in the field from various perspectives such as code smell types, deep learning techniques, datasets, and model performance evaluation. 
They highlighted supervised learning as the most commonly used learning method and pointed out the importance of models such as convolutional neural networks, recurrent neural networks, and long and short-term memory networks in CSD. In addition, they generally observed the prevalence of Java datasets and that method-level code smell is most often detected. 
Although these reviews discussed the datasets, there is a certain lack of detailed exploration and systematic analysis of the data preparation stages. They focused more on what type of code smell dataset was used, the programming language of the dataset, and the size of the dataset, while the impact of the dataset preparation process was not examined in depth. 

Compared to general CSD surveys, we specifically target the understudied domain of data preparation to advance understanding and inform future practices. There is one particular survey studying CSD datasets~\citep{zakeri2023systematic}. 
They meticulously compared CSD datasets across properties like size, supported smells, programming languages, and construction methods. Their findings highlighted several limitations within existing datasets, notably imbalances in samples, absence of severity levels for smells, and constraints related to Java-based datasets.
However, their analysis predominantly focused on machine learning datasets and did not explore the challenges identified in our survey, i.e., \textit{Data Scarcity}, \textit{Limited Generalization Ability}, \textit{Limited Data Accessibility}, \textit{Heavy Expert Dependency}, \textit{Difficulty of Data Labeling}, and \textit{Redundancy}.
In addition, they lacked a detailed exploration of solutions or recommendations to address the challenges found, particularly in the context of deep learning applications within CSD. 
Instead, we address these challenges by proposing diverse solutions.
Furthermore, we provide a set of recommendations to advance the field.
These recommendations aim to foster progress by addressing critical issues and promoting standardized practices within the domain of DL-based CSD.

\begin{figure*}[t]
    \includegraphics[width=1\linewidth]{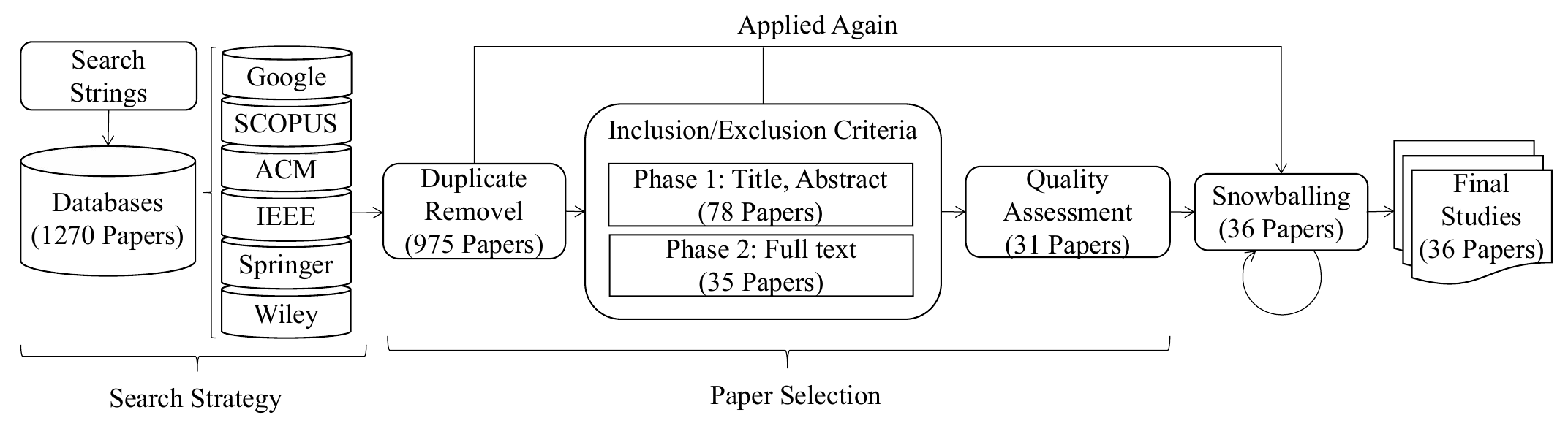}
    \caption{The overall process of our systematic literature review.}
    \label{SLR process}
\end{figure*}

\begin{table*}[t]
    \caption{The primary studies analyzed in our systematic literature review.}
    \begin{tabular}{p{0.5cm}p{2.7cm}p{11.8cm}}
    \toprule
    \textbf{ID} & \textbf{Reference} & \textbf{Title} \\ \midrule
    S1 & \scriptsize \citet{kim2017finding} & \scriptsize Finding Bad Code Smells with Neural Network Models \\ 
    S2 & \scriptsize \citet{hadj2018hybrid} & \scriptsize A Hybrid Approach To Detect Code Smells using Deep Learning \\ 
    S3 & \scriptsize \citet{fakhoury2018keep} & \scriptsize Keep It Simple: Is Deep Learning Good for Linguistic Smell Detection? \\ 
    S4 & \scriptsize \citet{guo2019deep} & \scriptsize Deep semantic-Based Feature Envy Identification \\ 
    S5 & \scriptsize \citet{barbez2019deep} & \scriptsize Deep Learning Anti-patterns from Code Metrics History \\ 
    S6 & \scriptsize \citet{liu2019deep} & \scriptsize Deep Learning Based Code Smell Detection \\ 
    S7 & \scriptsize \citet{hadj2019deep} & \scriptsize Deep Representation Learning for Code Smells Detection using Variational Auto-Encoder \\ 
    S8 & \scriptsize \citet{das2019detecting} & \scriptsize Detecting Code Smells using Deep Learning \\ 
    S9 & \scriptsize \citet{hadj2019improving} & \scriptsize Improving the Identification of Code Smells by Combining Structural and Semantic Information \\ 
    S10 & \scriptsize \citet{hamdy2020deep} & \scriptsize Deep Hybrid Features for Code Smells Detection \\  
    S11 & \scriptsize \citet{wang2020feature} & \scriptsize Feature Envy Detection based on Bi-LSTM with Self-Attention Mechanism \\ 
    S12 & \scriptsize \citet{yu2021novel} & \scriptsize A Novel Tree-based Neural Network for Android Code Smells Detection \\ 
    S13 & \scriptsize \citet{gupta2021empirical} & \scriptsize An Empirical Study on Predictability of Software Code Smell Using Deep Learning Models \\ 
    S14 & \scriptsize \citet{sharma2021code} & \scriptsize Code Smell Detection by Deep Direct-learning and Transfer-learning \\ 
    S15 & \scriptsize \citet{xu2021multi} & \scriptsize Multi-Granularity Code Smell Detection using Deep Learning Method based on Abstract Syntax Tree \\ 
    S16 & \scriptsize \citet{ren2021exploiting} & \scriptsize Exploiting Multi-aspect Interactions for God Class Detection with Dataset Fine-tuning \\ 
    S17 & \scriptsize \citet{yin2021local} & \scriptsize Local and Global Feature Based Explainable Feature Envy Detection \\ 
    S18 & \scriptsize \citet{ardimento2021temporal} & \scriptsize Temporal convolutional networks for just-in-time design \\  
    S19 & \scriptsize \citet{sidhu2022machine} & \scriptsize A machine learning approach to software model refactoring \\ 
    S20 & \scriptsize \citet{tarwani2022application} & \scriptsize Application of Deep Learning models for Code Smell Prediction \\ 
    S21 & \scriptsize \citet{khleel2022deep} & \scriptsize Deep convolutional neural network model for bad code smells detection based on oversampling method \\ 
    S22 & \scriptsize \citet{zhang2022delesmell} & \scriptsize Code smell detection based on deep learning and latent semantic analysis \\ 
    S23 & \scriptsize \citet{yedida2022improve} & \scriptsize How to Improve Deep Learning for Software Analytics \\ 
    S24 & \scriptsize \citet{li2022multi} & \scriptsize Multi-Label Code Smell Detection with Hybrid Model based on Deep Learning \\ 
    S25 & \scriptsize \citet{dewangan2022code} & \scriptsize Code Smell Detection Using Ensemble Machine Learning Algorithms \\ 
    S26 & \scriptsize \citet{zhang2022feature} & \scriptsize Feature Envy Detection with Deep Learning and Snapshot Ensemble \\ 
    S27 & \scriptsize \citet{bhave2022deep} & \scriptsize Deep Multimodal Architecture for Detection of Long Parameter List and Switch Statements using DistilBERT \\ 
    S28 & \scriptsize \citet{ardimento2021transfer} & \scriptsize Transfer Learning for Just-in-Time Design Smells Prediction using Temporal Convolutional Networks \\ 
    S29 & \scriptsize \citet{jeevanantham2022extension} & \scriptsize Extension of Deep Learning Based Feature Envy Detection for Misplaced Fields and Methods \\ 
    S30 & \scriptsize \citet{virmajoki2022detecting} & \scriptsize Detecting Code Smells with AI: a Prototype Study \\ 
    S31 & \scriptsize \citet{imam2022automation} & \scriptsize The automation of the detection of large class bad smell by using genetic algorithm and deep learning \\ 
    S32 & \scriptsize \citet{siddiq2022empirical} & \scriptsize An Empirical Study of Code Smells in Transformer-based Code Generation Techniques \\ 
    S33 & \scriptsize \citet{afrin2022hybrid} & \scriptsize A Hybrid Approach to Investigate Anti-pattern from Source Code \\ 
    S34 & \scriptsize \citet{ho2023fusion} & \scriptsize Fusion of deep convolutional and LSTM recurrent neural networks for automated detection of code smells \\ 
    S35 & \scriptsize \citet{kaur2023improving} & \scriptsize Improving the Quality of Open-Source Software \\ 
    S36 & \scriptsize \citet{liu2023deep} & \scriptsize Deep Learning Based Feature Envy Detection Boosted by Real-World Examples \\ \bottomrule
\end{tabular}
    \label{Papers}
\end{table*}

\begin{table*}[t]
    \caption{The key terms and synonyms for paper search. * denotes the wildcard matching pattern.}
    \begin{tabular}{p{1.5cm}p{3.2cm}p{10cm}}
    \toprule
        \textbf{Category} &\textbf{Subject} & \textbf{Search Terms} \\ \midrule
        Population & Code Smell & ``Bad smell*'' OR ``Design smell*'' OR ``Design flaw*'' OR ``Antipattern*'' OR ``Model smell*'' \\ 
        Intervention & DL Technique & ``Deep learning'' OR ``Transfer learning'' OR ``CNN'' OR ``RNN'' OR ``Auto-encoder*'' OR ``Deep neural network*'' \\
        Comparison & -& -\\
        Outcomes & Code Smell Detection & ``Detect*'' OR ``Predict*'' OR ``Identif*'' \\ 
        \bottomrule
    \end{tabular}
    \label{Seach String}
\end{table*}

\begin{table*}[t]
    \caption{The inclusion and exclusion criteria for screening eligible papers.}
    
    \begin{tabular}{ll}
    \toprule
    \textbf{Inclusion Criteria} & \textbf{Exclusion Criteria} \\ \midrule
    \textbf{IC1:} The paper proposes new deep learning techniques. & \textbf{EC1:} The paper is not in English.\\
    \textbf{IC2:} The paper reports on empirical results. & \textbf{EC2:} The paper is a literature review only. \\
    \textbf{IC3:} The paper has undergone peer review. & \textbf{EC3:} The full text of the paper is unavailable.\\
    & \textbf{EC4:} The paper provides no dataset(s) details.\\
    \bottomrule
    \end{tabular}
    \label{I/E Criteria}
\end{table*}

\section{Research Methodology} \label{SLR}
Our SLR process strictly adheres to established guidelines~\citep{kitchenham2004procedures, zhang2011identifying} to ensure an objective review. We also adopt a \textit{snowballing} approach~\citep{wohlin2014guidelines} to include additional literature and enhance the completeness of our review. Figure~\ref{SLR process} outlines our SLR process. The first two authors conduct the work closely with review from the other authors. This process occurred in 2023 and identified 36 relevant papers, as detailed in Table~\ref{Papers}.

\subsection{Search Strategy}
To design the search string for our SLR, we utilize the PICO (Population, Intervention, Comparison, Outcomes) framework~\citep{schardt2007utilization}. This framework is widely adopted in systematic reviews to formulate research questions and develop search strategies. PICO helps in breaking down the research topic into four key components:

\begin{itemize}
    \setlength{\itemsep}{0pt}
    \setlength{\parsep}{0pt}
    \setlength{\parskip}{0pt}
    \item Population (P): The population of interest in our study is represented by ``Code Smell''.
    \item Intervention (I): The intervention refers to the ``DL Technique'' (Deep Learning Technique).
    \item Comparison (C): Comparison is not applicable in our study, hence it is omitted.
    \item Outcomes (O): The expected outcomes are related to ``Code Smell Detection''.
\end{itemize}

Table \ref{Seach String} details the key terms and synonyms associated with each PICO component used in our search strategy. We constructed the query by combining these PICO components using the Boolean operator ``AND''. This approach ensures a comprehensive search by encompassing a broad spectrum of research related to our topic. We include variants of key terms facilitated through the use of wildcard matching. For instance, the term ``detect*'' covers ``detect'', ``detection'', ``detecting'', etc. We combine the key terms using the ``OR'' operator. The detailed search string in the SCOPUS format is as follows:

\textit{TITLE-ABS-KEY((``Code smell'' OR ``Bad smell*'' OR ``Design smell*'' OR ``Design flaw*'' OR ``Antipattern*'' OR ``Model smell*'') AND (``DL Technique'' OR ``Deep learning'' OR ``Transfer learning'' OR ``CNN'' OR ``RNN'' OR ``Auto-encoder*'' OR ``Deep neural network*'') AND (``Code Smell Detection'' OR ``Detect*'' OR ``Predict*'' OR ``Identif*'')}

We adapt the search string as necessary to match each database. The databases queried include Google Scholar, SCOPUS, ACM Digital Library, IEEE Xplore, Springer, and Wiley. These databases are chosen based on recommendations from previous SLRs~\citep{yang2022survey, martinez2022software}, which highlighted them as sources containing high-quality, peer-reviewed research in software engineering.

Furthermore, we apply additional filters to the retrieved papers, including the language (i.e., English-only) and publication status (i.e., The paper should be a peer-reviewed full research paper published in a conference proceeding or a journal). The initial search identifies 1270 papers. We then remove the duplicate records, resulting in 975 papers for further screening.

\subsection{Paper Selection}
We aim to identify high-quality studies that could provide valuable insights into data preparation for DL-based CSD. Additional criteria and quality assessment are applied to screen eligible papers.

\subsubsection{Inclusion/Exclusion Criteria}
We propose three inclusion and four exclusion criteria, as shown in Table~\ref{I/E Criteria}.  A paper is only included if it meets all the inclusion criteria and does not conform to any exclusion criteria. The inclusion criteria require that papers utilize deep learning techniques for CSD and propose novel deep learning-based models or solutions for the task. Papers also need to describe the datasets used clearly. For exclusion, papers that use heuristic or machine learning-based detection techniques, review existing models, or only perform statistical/correlational analyses are removed. Papers with unclear descriptions of the datasets are also excluded. To help validate the consistent application of the criteria, the first two authors also conduct an initial screening of 50 randomly selected papers. They independently assess whether each paper meets or does not meet the inclusion and exclusion criteria. 
The absence of discrepancies between the authors' assessments lends additional confidence in the reliability and validity of the criteria used in this study.

The criteria are applied in two phases. First, the titles and abstracts of retrieved papers are screened according to the criteria. This process leaves 78 papers for potential inclusion. Then, the full texts of the remaining 78 papers are thoroughly reviewed against the criteria. After a full assessment, we have 35 papers that suit the inclusion and exclusion criteria.

\subsubsection{Quality Assessment}
\begin{table*}[t]
    \caption{The quality criteria checklist for screening eligible papers.}
    
    \begin{tabular}{l}
    \toprule
    \textbf{Quality Criteria} \\ \midrule
    \textbf{QC1:} Are the code smells being detected clearly defined?\\
    \textbf{QC2:} Are the deep learning models sufficiently described?\\
    \textbf{QC3:} Are the performance metrics specified?\\
    \textbf{QC4:} Are the independent and dependent variables clearly defined?\\
    \textbf{QC5:} Are the data sources and statistics fully described?\\
    \textbf{QC6:} Is the data labeling method clearly explained?\\
    \textbf{QC7:} Is the validation methodology specified?\\
    \textbf{QC8:} Are potential threats to validity clearly outlined?\\
    \bottomrule
    \end{tabular}
    \label{Quality Checklist}
\end{table*}

\begin{table*}[t]
    \caption{The data extraction form for collecting information from reviewed papers.}
    
    \begin{tabular}{p{2cm}p{4cm}p{6cm}}
    \toprule
    & \textbf{Item} & \textbf{Description} \\ \midrule
    \multirow{6}{*}{Metadata} & Study ID & Unique identifier for the paper \\
    ~ & Title & Title of the paper\\
    ~ & Author & Author(s) of the paper\\
    ~ & Year & Year of publication\\
    ~ & References & Number of references in the paper\\
    ~ & Publication & Journal or conference of publication\\ \midrule
    \multirow{10}{*}{Datasets} & Data Source & Real-world or synthetic data\\
    ~ & Dataset Name & Name of dataset(s) used\\
    ~ & Multiple Datasets & Number of datasets for experiments\\
    ~ & Data Integration & How multiple datasets were used\\
    ~ & Availability & Reproducibility of datasets\\
    ~ & Source Type & Open-source or exclusive license\\
    ~ & Code Smell Types & Types of code smells considered\\
    ~ & Programming Language & Language of code in datasets\\
    ~ & Size & Number of samples in datasets\\
    ~ & Ratio & Ratio of smelly to non-smelly samples\\
    ~ & Labeling Method & Approach to labeling smelly samples\\
    ~ & Required Expertise & Expertise needed for labeling\\ \midrule
    \multirow{5}{*}{Methods} & Data Cleaning & Pre-processing approaches\\
    ~ & Transformation & Data representation techniques\\
    ~ & Partitioning & How data was split for training/evaluation\\
    ~ & Resampling & Methods to handle class imbalance\\
    ~ & DL Techniques & Proposed deep learning approaches\\ \midrule
    ~ Others & & Additional relevant findings\\
    \bottomrule
    \end{tabular}
    \label{Data Extraction}
\end{table*}

We conduct a quality assessment on the remaining papers using the checklist in Table~\ref{Quality Checklist}. Quality criteria are essential for assessing the reliability of extracted information, though there is no standardized approach~\citep{kitchenham2009systematic}. Following the \citet{alazba2023deep,croft2022data}, and \citet{zakeri2023systematic}, our checklist mainly examines independent/dependent variables, validation methods, datasets, and experimental complexity.

One author initially performs the quality assessment, with two additional authors conducting another round of results validation. This process excludes four papers for failing to meet one or more quality criteria. Specifically, \citet{lin2021novel}, \citet{virmajoki2020detecting}, \citet{malathi2023class}, and \citet{grodniyomchai2019deep} lack sufficient descriptions of independent/dependent variables, validation approaches, or datasets used. The details of these papers that do not meet our quality standards are omitted for brevity. This assessment aims to screen papers with incomplete reporting that could limit the extraction of meaningful insights.

\subsection{Snowballing}
To ensure comprehensive coverage of all relevant literature, we perform manual \textit{snowballing} as per the guidelines by Wohlin et al.~\citep{wohlin2014guidelines}. This involves both forward and backward snowballing techniques. Forward snowballing involves examining the citation lists of all papers that meet our inclusion criteria to locate additional relevant papers. Backward snowballing reviews the reference lists to uncover any pertinent studies not previously identified.
During the initial round of snowballing, we successfully identify five new relevant papers. Subsequent rounds of snowballing are conducted following the same rigorous screening process, adhering to our predefined inclusion and exclusion criteria and maintaining our quality assessment standards.  Despite the additional rounds, no new papers are found that meet our criteria, leading us to conclude that we have reached a saturation point. This is due to the limited scope of current research on DL-based code smell detection, which is a relatively nascent field.

In total, our search and snowballing processes yield 36 papers. Of these, 31 papers are initially retrieved through keyword searches across various databases. The remaining five papers are found through manual snowballing of references and citations. This dual-phase approach helps provide a more comprehensive examination of the literature.  
\begin{table*}[!ht]
    \centering
    \caption{The publication statistics of the primary studies in our SLR. \textit{J} and \textit{C} denote journal and conference publication, respectively. \textit{N.} denotes the number of publications.}
    \begin{tabular}{p{13cm}p{0.5cm}p{1.5cm}}
    \toprule
        \textbf{Sources} & \textbf{N.} & \textbf{Study} \\ \midrule
        J: IEEE Transactions on Software Engineering & 1 & S[6] \\ 
        J: Journal of Systems and Software & 1 & S[14] \\ 
        J: Neurocomputing & 1 & S[18] \\ 
        J: Knowledge-Based Systems & 1 & S[22] \\ 
        J: International Journal of Electrical and Computer Engineering & 1 & S[1] \\ 
        J: Journal of Theoretical and Applied Information Technology & 1 & S[10] \\ 
        J: International Journal of Computers and Applications & 1 & S[19] \\ 
        J: Indonesian Journal of Electrical Engineering and Computer Science & 1 & S[21] \\ 
        J: Applied Sciences & 1 & S[25] \\ 
        J: International Journal of Intelligent Engineering and Systems & 1 & S[29] \\ 
        J: Journal of King Saud University-Computer and Information Sciences & 1 & S[31] \\ 
        J: Agile Software Development: Trends, Challenges and Applications & 1 & S[35] \\ 
        C: International Conference on Software Engineering and Knowledge Engineering & 2 & S[15, 24] \\ 
        C: International Computer Software and Applications Conference & 2 & S[16, 17] \\ 
        C: IEEE International Working Conference on Source Code Analysis and Manipulation & 2 & S[27, 32] \\ 
        C: ACM Joint European Software Engineering Conference and Symposium on the Foundations of Software Engineering & 1 & S[36] \\ 
        C: International Conference on Software Maintenance and Evolution & 1 & S[5] \\ 
        C: International Conference on Software Quality, Reliability and Security & 1 & S[12] \\ 
        C: International Conference on Neural Information Processing & 1 & S[9] \\ 
        C: International Conference on Evaluation and Assessment in Software Engineering & 1 & S[34] \\ 
        C: IEEE International Symposium on Parallel and Distributed Processing with Applications & 1 & S[11] \\ 
        C: Asia-Pacific Symposium on Internetware & 1 & S[4] \\ 
        C: International Conference on Evaluation of Novel Approaches to Software Engineering & 1 & S[2] \\ 
        C: International Conference on Software Analysis, Evolution and Reengineering & 1 & S[3] \\ 
        C: International Joint Conference on Neural Networks & 1 & S[7] \\ 
        C: IEEE Region 10 Conference & 1 & S[8] \\ 
        C: International Conference on Advanced Information Networking and Applications & 1 & S[13] \\ 
        C: International Conference on Reliability, Infocom Technologies and Optimization (Trends and Future Directions) & 1 & S[20] \\ 
        C: International Conference on Mining Software Repositories & 1 & S[23] \\ 
        C: International Conference on Dependable Systems and Their Applications & 1 & S[26] \\ 
        C: Jubilee International Convention on Information, Communication and Electronic Technology & 1 & S[30] \\ 
        C: International Conference on Computer and Information Technology  & 1 & S[33] \\ 
        C: International Conference on Software Technologies & 1 & S[28] \\ \bottomrule
    \end{tabular}
    \label{Publication Sources}
\end{table*}

\begin{figure}
 \centering
 \includegraphics[width=0.45\linewidth]{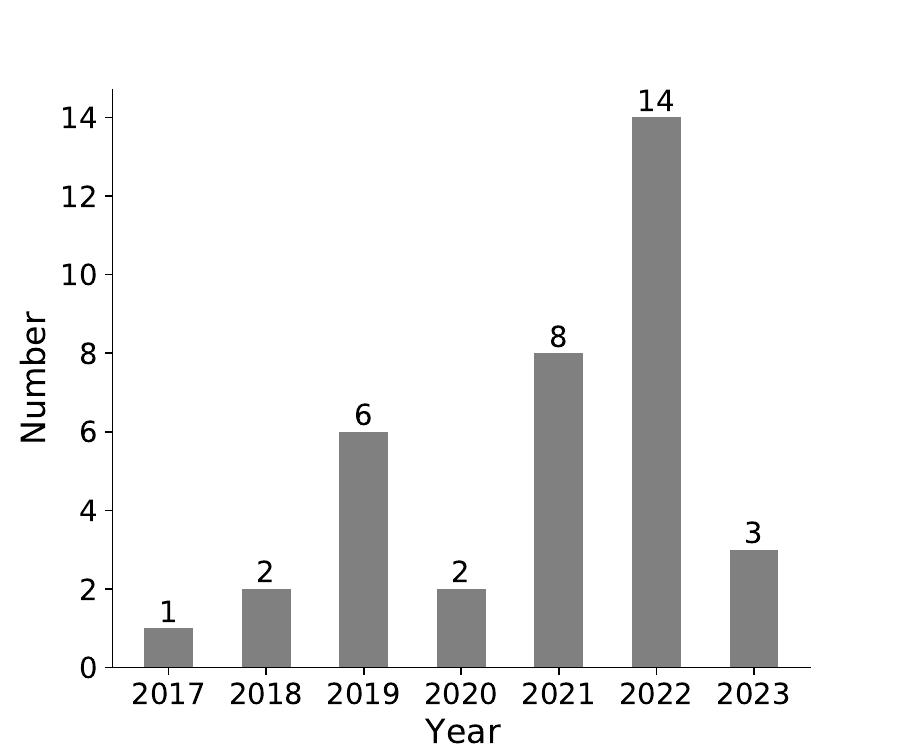}
 \caption{The number of primary studies by year.}
 \label{Year}
\end{figure}

\subsection{Data Extraction}
We have designed a data extraction form to systematically analyze the identified papers, shown in Table~\ref{Data Extraction}. The form design is adapted from prior SLR guidelines~\citep{garousi2017experience, kitchenham2004procedures} and pilot-tested before finalizing.

The form captures qualitative and quantitative attributes across four aspects: metadata, datasets, methods, and other information. One author performs the initial data extraction to organize information collected from each paper. Then, two additional authors verify the extracted data through independent examination. Any disagreements are resolved through group discussion to reach a consensus.
Most data extracted is qualitative, such as deep learning techniques applied, data sources, pre-processing approaches, and code smells addressed. Some quantitative data is also collected, like imbalance ratios within datasets. This formal extraction process aims to investigate the research questions proposed in our study comprehensively.  
The publication details of the 36 primary studies are listed in Table~\ref{Publication Sources}.  
The number of these 36 papers over the years is shown in Figure~\ref{Year}.

\section{RQ1 - Critical Considerations in CSD Data Preparation} \label{RQ1}

\begin{figure*}
 \centering
 \includegraphics[width=0.9\linewidth]{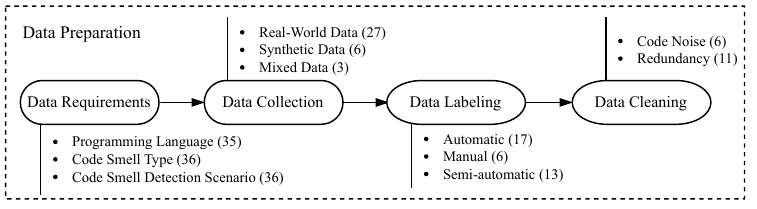}
  \caption{The critical considerations in CSD data preparation (RQ1).The number of papers for each category is indicated.}
 \label{Fig_RQ1}
\end{figure*}

Through a comprehensive literature review, we extract and analyze the various considerations taken to construct datasets for code smell detection. Following the machine learning workflow introduced by ~\citet{amershi2019software}, our data preparation analysis centers around four main phases in Figure~\ref{Fig_RQ1}, including data requirements, collection, labeling, and cleaning. 
By thoroughly reviewing these preparation aspects, we clarify current practices and guide practitioners and researchers on effectively addressing critical factors when building datasets. This will help standardize the construction of high-quality datasets for code smell detection.

\subsection{Data Requirements}
When constructing high-quality datasets to train and evaluate DL-based code smell detection models, it is crucial to determine which programming language code will undergo smell detection and the specific types of code smells to be detected. Moreover, we should also consider the code smell detection scenario, i.e., whether to use within-project or cross-project data to build the datasets. Therefore, three key factors should be considered when preparing datasets for code smell detection research: programming language, code smell type, and code smell detection scenario.

\begin{figure}
 \centering
 \includegraphics[width=0.3\linewidth]{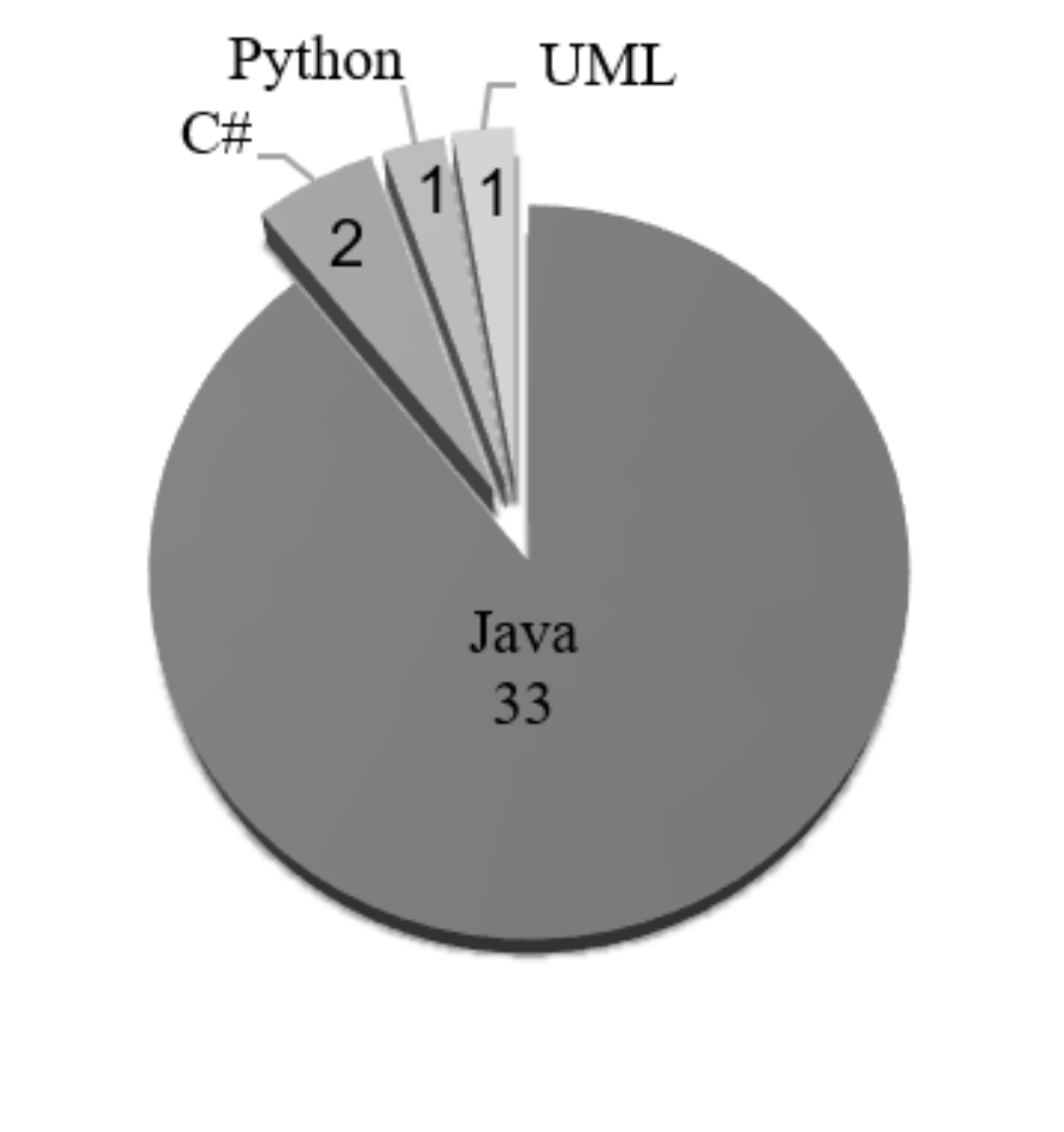}
 \caption{The frequency of programming languages addressed in analyzed primary studies.}
 \label{Programming Language}
\end{figure}

\paragraph{Programming Language:} 
The choice of programming language is an essential early decision in dataset construction. Several aspects influence this choice, including the availability of openly accessible code samples and the types of code smells to be studied for that particular language. As depicted in Figure \ref{Programming Language}, our analysis shows that the vast majority of papers [S1-12, S14-18, S20-31, S33-36] utilize Java datasets due to the widespread use of the Qualitas Corpus --- an open-source collection of Java projects. The higher availability of Java datasets helps accelerate research in this area.
Besides, two papers [S14, S34] focus on C\# to investigate the feasibility of transfer learning across languages. 
There is one paper [S32] studying Python datasets and one paper [S19] studying UML datasets, where [S32] specifically examines Python security smells, and [S19] studies the presence of functional decomposition in UML models of object-oriented software. S13 does not provide details on the programming language studied or the dataset used, which is not categorized within this section.

\paragraph{Code Smell Type:} 
Another critical consideration is the types of code smells.
In our study, we categorize the code smells into the \textit{Class} level, the \textit{Method} level, and the code smells that are relevant to \textit{Both} levels. 
The \textit{Class} level code smells typically pertain to the design and structure of entire classes, which concerns class-level refactoring or redesign. 
The \textit{Method} level code smells are usually related to the internal implementation and behavior of the methods or functions, which concerns refactoring or decomposition of the methods. 
The \textit{Both} level represents code smells that may result in both class- and function-level refactoring.
We list the details of the \textit{Method} level code smells, the \textit{Class} level, and the \textit{Both} level code smells in Table \ref{Code Smell Type} and \ref{Code Smell Type Continued}.

As can be seen, certain code smell types have garnered considerable attention from researchers, with four prevalent types covered by ten or more papers. Among them, \textit{Feature Envy} is the most investigated
code smell, which denotes methods that access the data of another object rather than its data. Another prevalent code smell \textit{God Class} \footnote{Also known as the \textit{Blob Class}, which is studied in [S7, S9, S13, S33].}  means classes that have many members and implement different behaviors. The \textit{Long Method} code smell refers to the methods that are too long and contain too much code logic. And \textit{Data Class} means a class containing only data fields and methods. 
The prevalence of these code smell types can be attributed to their distinctiveness, ease of identification, and relatively large number of samples in real-world codebases. 

However, some code smell types have not received enough attention. One of them is \textit{Type checking}, which means frequently using type checking to determine the type of an object. Another example is \textit{Dummy handler}, an exception handler that only logs an error message without taking any meaningful corrective actions. Notably, there has been a recent effort~\citep{tarwani2022application} studying these code smells to broaden the scope of experimental research and enhance empirical investigations in these domains.

\begin{table*}
\centering
    \caption{The code smell types addressed in analyzed studies.}
    \begin{tabular}{p{3.6cm}p{8.2cm}p{3cm}}
\toprule
    \textbf{Code Smell Type} & \textbf{Description} & \textbf{Study} \\ \midrule
    \rowcolor{lightgray} \multicolumn{3}{c}{\textit{Method Level}}\\
        Feature Envy & Methods accessing another object's data. & S[1-2, 4, 6-7, 9, 11, 14-15, 17, 20-21, 23, 25-26, 29-30, 34-36] \\ 
        Long Method & Methods with excessive code logic. & S[2, 6-7, 9, 13, 20-21, 23-25, 30] \\ 
        Complex Method & Methods with high cyclomatic complexity. & S[14, 24, 34] \\ 
        Empty Catch Block & Empty exception catch blocks. & S[15, 20, 24] \\ 
        Brain Method & Methods concentrating class intelligence excessively. & S[8, 22, 35] \\ 
        Complex Conditional &  Long or complex conditional expressions. & S[14, 24, 34] \\ 
        Member Ignoring Method & Ordinary methods not accessing member attributes. & S[12-13] \\ 
        Internal Getter and Setter & Methods accessing properties via get/setters. & S[12-13] \\ 
        Long Parameter Lists & Methods with excessively long parameter lists. & S[24, 27] \\ 
        Magic Number & Unexplained numeric literals in expressions. & S[24] \\ 
        Type checking & Frequent object type checking. & S[20] \\ 
        Shotgun Surgery & Similar changes in multiple places for requirements. & S[22] \\ 
        Over logging & Excessive logging causing large log files. & S[20] \\ 
        No Low Memory Resolver & Lack of proper low memory handling. & S[13] \\ 
        Nested try statement & Multiple layers of nested try-catch blocks. & S[20] \\ 
        Linguistic Antipatterns & Poor language in code, comments, or documentation. & S[3] \\ 
        Exception in finally block & Throwing exceptions within a finally block. & S[20] \\ 
        Dummy handler & Handling exceptions by just printing error messages. & S[20] \\ 
        Careles Cleanup & Mishandling exceptions or resource leaks in cleanup. & S[20] \\ 
        SpaghettiCode & Confusing and intricate code structure. & S[33] \\ 
        Intensive Coupling & Methods calling too many other member methods. & S[35] \\ 
        Extensive Coupling & Methods calling scattered member methods. & S[35] \\ 
        Switch Statements & Heavy use of switch statements. & S[27] \\ 
        Long Identifier & Excessively long identifiers. & S[24] \\ 
        Long Statement & Individually lengthy statements. & S[24] \\ 
        Missing default & Lack of a default case branch in switch statements. & S[24] \\ \midrule
        \end{tabular}
    \label{Code Smell Type}
\end{table*}

\begin{table*}
\centering
    \caption{The code smell types addressed in analyzed studies (Continued).}
    \begin{tabular}{p{4.4cm}p{7.5cm}p{3cm}}
\toprule
    \textbf{Code Smell Type} & \textbf{Description} & \textbf{Study} \\ \midrule
\rowcolor{lightgray} \multicolumn{3}{c}{\textit{Class Level}}\\
        God Class (Blob Class) & Classes with numerous behaviors. & S[1-2, 5, 7, 9, 10, 13, 16, 20-21, 25, 33, 35] \\ 
        Data Class & Classes containing only fields and access methods. & S[1-2, 21-22, 25, 35] \\ 
        Large Class & Classes with many methods and data members. & S[1, 6, 23, 31] \\ 
        Multifaceted Abstraction & Classes having multiple responsibilities. & S[14, 34] \\ 
        Misplaced Class & Classes improperly distributed. & S[6, 23] \\ 
        Leaking Inner Class & Inner classes referencing outer classes. & S[12-13] \\ 
        Brain Class & Overly complex classes. & S[8, 22] \\ 
        Swiss Army Knife & Classes using multiple interfaces for functionalities.& S[13, 33] \\ 
        Functional Decomposition & Class functionality spread across multiple classes. & S[19, 33] \\ 
        Unprotected main & Core logic in an unprotected main function. & S[20] \\ 
        Parallel Inheritance Hierarchies & Inheritance tree dependencies. & S[1] \\ 
        Lazy Class & Classes not performing enough. & S[1] \\ 
        Insufficient Modularization & Incomplete class decomposition. & S[15] \\ 
        Deficient Encapsulation & Over-permissive member accessibility. & S[15] \\ 
        Complex Class & Classes with intricate logic. & S[13] \\ 
        Schizophrenic Class & Classes with unrelated functions. & S[35] \\ 
        Refused Parent Bequest & Subclasses resisting parent class methods. & S[35] \\ \midrule
        \rowcolor{lightgray} \multicolumn{3}{c}{\textit{Both Level}}\\
        Design Smell & Poor design choices in software systems. & S[18, 28] \\ 
        Security Smells & Potential security holes or vulnerabilities in the code. & S[32] \\ \bottomrule
 \end{tabular}
    \label{Code Smell Type Continued}
\end{table*}

\begin{figure}
 \centering
 \includegraphics[width=0.5\linewidth]{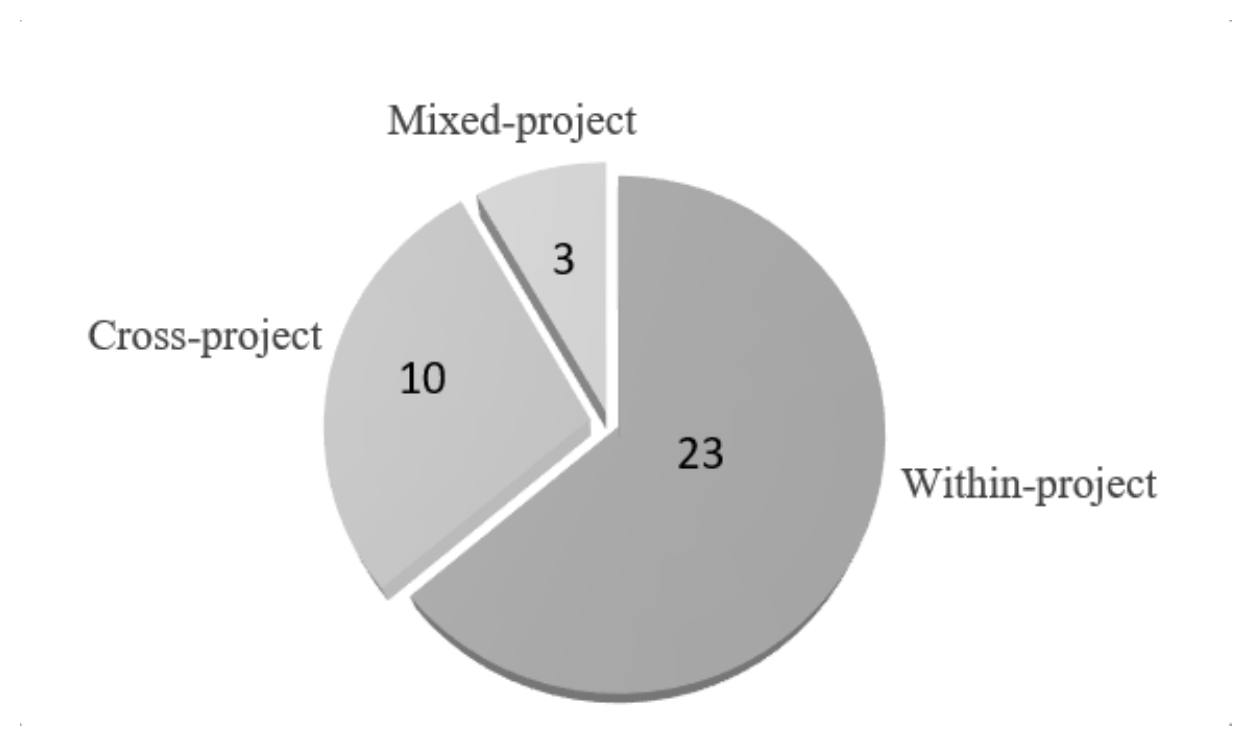}
 \caption{The distribution of code smell detection scenarios in analyzed primary studies.}
 \label{Application Contexts}
\end{figure}

\paragraph{Code Smell Detection Scenario:}
There are three main CSD scenarios. 
First is within-project detection, splitting a project into the training and testing data with no intersection. The second is cross-project detection. Contrary to the previous, the training and testing data come from different projects. This approach solves the problem of lacking enough training data from a single project. The third is mixed-project detection, which utilizes mixed data from multiple projects to get the training and testing data partitions. In this way, it can create enough data for training and evaluation.
We categorize the reviewed papers based on their CSD scenario and draw a pie chart in Figure \ref{Application Contexts}. We can find that 23 papers [S1-4, S7, S9, S12-13, S15, S17-23, S25-27, S29-30, S32-33] belong to the within-project scenario; Ten papers [S5-6, S8, S11, S14, S16, S28, S34-36] belong to the cross-project scenario; And three papers [S10, S24, S31] belong to the mixed-project scenario.
Within-project detection is the most popular practice for training and testing CSD models. There is a scarcity of papers using mixed-project datasets because unifying the feature extraction from different projects is still an open challenge.

\subsection{Data Collection}
The primary considerations during data collection vary based on the data source, which we categorize as real-world, synthetic, or mixed.

\paragraph{Real-World Data:}
Most studies [S1-5, S7-9, S12-15, S18-22, S24-25, S27-28, S30, S32-36] utilize real-world data by collecting open-source projects/repositories or using existing datasets. Real-world data is the best testbed for validating CSD techniques in practical applications. The most commonly used dataset is Qualitas~\citep{tempero2010qualitas}, which contains many Java open-source projects. It is used by five papers [S2, S4, S21, S25, S27]. Other well-processed corpora are constructed using multi-lingual source code from Github, Bitbucket, Apache, etc. Examples include  LandFill~\citep{palomba2015landfill} [S7, S9],  MUSE~\citep{yu2021novel} [S12],  MLCQ~\citep{madeyski2020mlcq} [S30], CodeXGlue~\citep{lu2021codexglue} [S32], and the Benchmark~\citep{sharma2021code} [S14, S34]. Furthermore, there is also 
a study [S19] investigating alternative data type, using the Img2UML~\citep{karasneh2013img2uml} corpus, which consists of XMI file of the UML class models parsed from images. Utilizing these real-world datasets is crucial in CSD research as they provide valuable insights into the complexities and challenges faced in practical software development.

\paragraph{Synthetic Data:}
In the context of CSD, researchers need to generate synthetic data to tackle challenges like insufficient real-world code smell samples or severe data imbalance. We identify six papers [S6, S11, S17, S23, S26, S29] using synthetic methods to overcome these challenges.
They usually follow a unified process for synthesizing the needed data. The first is to collect usable code snippets. The second is to assess whether each code snippet can be transformed into a code smell. The final step is to generate positive and negative samples using the identified code snippets in the second step. Negative samples are the unchanged code snippets. 
Positive samples are artificially altered fragments of the original code to make it smelly. For example, to create feature envy smells, they can perform unnecessary move refactoring, moving methods from one class to another~\cite{liu2019deep}.

\paragraph{Mixed Data:}
Another way to create datasets is to mix real-world and synthetic data. This is typically employed to address the challenge of having limited samples while preserving real-world data distribution~\citep{di2018detecting}. 
Our survey identifies three papers [S10, S16, S31] that use mixed data. Specifically, [S10] mixes the real-world data from the Qualitas and synthetic data. [S16] mixes synthetic data created by \citet{liu2019deep} with real-world data from the LandFill~\citep{ren2021exploiting}. [S31] mixes real-world~\citep{arcelli2016comparing,sousa2017findsmells} and synthetic datasets~\citep{liu2019deep} from previous references used.
These mixed datasets provide researchers with a valuable resource for conducting experiments that balance real-world complexity's benefits with synthetic data's controlled environment.

\subsection{Data labeling} \label{Data labeling}
The scale and quality of data labeling significantly influence the results and reliability of empirical studies. We summarize three labeling methods for constructing CSD datasets: automatic, manual, and semi-automatic ways.

\paragraph{Automatic:}
A common practice of labeling CSD datasets is using automatic tools. Such practice can bring several benefits, including convenience and time efficiency~\citep{liu2019deep}. 17 papers [S4-6, S8, S10-13, S15, S21-24, S26, S29-30] use automatic tools to label the datasets.
One of the frequently used tools is JDeodorant~\citep{tsantalis2008jdeodorant}. It is an Eclipse plug-in that detects code smells in Java software and recommends appropriate refactorings to resolve them. For the moment, the tool supports five code smells, namely \textit{Feature Envy}, \textit{Type/State Checking}, \textit{Long Method}, \textit{God Class}, and \textit{Duplicated Code}. Another example is Checkstyle~\citep{sourceforgeCheckstylex2013}, which is a development tool for Java, which checks many aspects of the source code. It can find class and method design problems. It also has the ability to check code layout and formatting issues. 
We provide a summary of all identified automatic tools in Table~\ref{Tool_Aotu}.

\begin{table*}[!ht]
    \centering
    \caption{The summary of the automatic tool for code smell detection.}
    \begin{tabular}{p{2cm}p{9cm}p{4cm}}
    \toprule
        \textbf{Tool} & \textbf{Code Smell} & \textbf{Study}\\ \midrule
        JDeodorant~\citep{tsantalis2008jdeodorant} & Feature Envy, Type/State Checking, Long Method, God Class, Duplicated Code & S[5-6, 11-12, 16, 18, 22, 26, 28, 29, 36] \\ 
        iPlasma~\citep{marinescu2005iplasma} & Duplicated Code, God Class, Feature Envy, Refused Bequest & S[2, 4, 8, 10, 16, 21, 22, 25] \\ 
        PMD~\citep{pmd} & Large Class, Long Method, Long Parameter List, Duplicated Code & S[2, 10, 12, 21, 25, 28] \\ 
        AntiPattern~\citep{wieman2011anti} & Data Class, Feature Envy, Long Method & S[2, 10, 21, 25] \\ 
        Checkstyle~\citep{sourceforgeCheckstylex2013} & Large Class, Long Method, Long Parameter List, Duplicated Code & S[12] \\ 
        UCDetector~\citep{ucdetector} & Data Class, Large Class, Long Method, Long Parameter List, Message Chains, Refused Bequest, Speculative Generality, Tradition Breaker & S[12] \\ \bottomrule
    \end{tabular}
    \label{Tool_Aotu}
\end{table*}

\paragraph{Manual:}
Manual labeling is time-consuming and labor-intensive, demanding substantial human resources and specialized knowledge. However, manual effort is sometimes necessary because humans can easily generalize to different domains and achieve higher reliability. We identify six papers [S1, S7, S9, S17, S19-20] that manually label the datasets. The manual labeling process first requires experts to manually analyze the source code based on various code smells. Secondly, additional experts are required to validate the accuracy of the previously identified smelly samples. Any disputed samples should be reviewed again with respect to code smell definitions, source code, and change history information to reach a conclusion. Such protocol enhances the reliability and reduces the subjectivity of the labeling process ~\cite{palomba2015landfill}.

\paragraph{Semi-automatic:}
There are 13 papers [S2-3, S14, S16, S18, S25, S27, S31-36] that explore a hybrid approach that combines both methods to address the reliability issue associated with automatic labeling and the labor-intensive nature of manual labeling.
Generally, they use automatic tools to label all samples and verify the labeled results by experts. Specifically, a subset of samples regarding code smells is chosen for individual analysis by multiple experts. The experts perform individual analyses without discussing them with others. Then, Cohen's Kappa is calculated to measure inter-expert agreement. Disagreements are discussed to reach a consensus on a final labeled set. Finally, the manually labeled results are compared to the automatic tools' labeled results. If the differences are negligible, the dataset labeled by the automatic tools for all samples is ultimately adopted.

\subsection{Data Cleaning}
The final stage of data preparation is data cleaning. Though not all studies comprehensively address this stage, we identify two prevalent cleaning steps involving code noise and data redundancy.

\paragraph{Code Noise:}
Six papers [S3, S18, S21, S27-28, S33] indicate that code noise may introduce irrelevant or erroneous information that can mislead models. [S21, S27-28] identifies several noise types, including outliers, missing data, and mismatching feature types. Textual noise like blank lines and non-ASCII characters are also found in [S3]. [S18, S27] find that incomplete or erroneous data sessions and non-normalized features can introduce additional noise. Object data types instead of expected numerical types are also identified in one dataset [S33]. Such noise can be removed through pre-processing to improve dataset usability for CSD models, which will be detailed in the subsequent sections of this survey. 

\paragraph{Redundancy:}
Data redundancy refers to identical or highly similar code samples and redundancy features. This adversely impacts analysis and model performance. To mitigate these effects, two papers [S14-15] remove duplicate code samples from identical code files or fragments. 
Nine papers [S6, S9-10, S13, S18-21, S25] use various feature selection methods to remove redundant features, including:

\begin{itemize}
    \setlength{\itemsep}{0pt}
    \setlength{\parsep}{0pt}
    \setlength{\parskip}{0pt}
    \item Convolutional Neural Networks (CNN): Applied in [S6, S10, S18, S21], CNNs are leveraged for their ability to automatically identify and discard redundant features through generalizing from relevant data.
    \item Gain Ratio: Utilized in [S9, S21], the gain ratio is an extension of the information gain criterion, which normalizes the information gain by the intrinsic information of a split, making it effective in choosing features that provide the most significant discrimination.
    \item Cross-Correlation Analysis: As described in \citep{podobnik2007detrended} and used in [S13], this method involves analyzing the cross-correlation function to identify and eliminate features that exhibit high redundancy with other features.
    \item Chi-Square Test: Employed in [S19, S25], the chi-square test evaluates the independence of features with respect to the target variable, enabling the selection of features that have a statistically significant association with the outcome, thus removing irrelevant or redundant features.
    \item Information Gain: Applied in [S20], information gain measures the reduction in entropy or uncertainty by partitioning the data according to different features, helping to identify and retain the most informative features.
\end{itemize}

\section{RQ2 - Challenges in CSD Data Preparation} \label{RQ2}
This section presents seven prominent challenges encountered by researchers during the creation of CSD datasets, including \textit{Data Scarcity}, \textit{Limited Generalization Ability}, \textit{Limited Data Accessibility}, \textit{Heavy Expert Dependency}, \textit{Difficulty of Data Labeling}, \textit{Data Imbalance}, and \textit{Data Redundancy}. Each challenge presents a unique hurdle in the quest for effective DL-based CSD. We briefly summarize these challenges in Table \ref{RQ2 summary}.

\begin{table*}[ht!]
    \centering
    \caption{The summary of data challenge in CSD.}
    \begin{tabular}{p{4.1cm}p{7cm}p{4cm}}
    \toprule
        \textbf{Challenge} & \textbf{Description} & \textbf{Study} \\ \midrule
        Data Scarcity & · Lack of large, real-world datasets to train DL models. & S[3, 5-6, 14, 16, 22, 36] \\ 
        ~ & · Synthetic data cannot represent real code smells.  & S[6, 11, 16-17] \\
        Limited Generalization Ability & · A single programming language during training hinders model generalization. & S[1, 3, 7, 12, 15, 22, 24-25, 28, 30, 32, 36] \\ 
        ~ & · Including only a few code smell types in datasets restricts models from detecting other smell types. & S[1, 5-6, 9, 12, 15-18, 22, 25, 30, 32, 35] \\ 
        ~ & · Using datasets solely from open-source projects limits generalizing to proprietary codebases.  & S[2, 7, 15, 18, 28, 35] \\ 
        Limited Data Accessibility & · Lack of clarity about dataset construction makes it difficult to reproduce. & S[3, 8, 13-15, 22, 24, 28-30] \\ 
        ~ & · Using private datasets limits independent validation and extension of approaches. & S[6] \\ 
        Heavy Expert Dependency & · Manual labeling by domain experts is crucial but demanding given the scale of needed datasets. & S[3, 6, 14, 17-18, 26, 34, 36] \\ 
        Difficulty of Data Labeling & · Labeling is time-consuming and error-prone, with challenges in accuracy for automated labeling and potential noise in human labels.  & S[5-6, 18, 24, 36] \\ 
        Data Imbalance & · Uneven distribution of code smell samples hinders model training. & S[4, 12-16, 21-23, 25, 27-28, 31, 33] \\ 
        Redundancy & · Duplicate samples inflate dataset sizes without adding value for learning. & S[2, 14-15] \\ 
        ~ & · Redundant and uninformative features makes model training more difficult. & S[2, 13, 21, 27] \\ \bottomrule
    \end{tabular}
    \label{RQ2 summary}
\end{table*}

\subsection{Data Scarcity}
Data scarcity in DL-based CSD refers to the inadequacy of available real-world data for training and testing DL models. Seven papers [S3, S5-6, S14, S16, S22, S36] point out this issue. In cases where the datasets lack sufficient samples, the model may struggle to acquire essential features and patterns, ultimately leading to a decline in performance~ \citep{fakhoury2018keep,sharma2021code}. For example, [S5] states that their sample size may limit the generalizability of the results and hope further to evaluate the approach on a larger set of systems. 
In addition, several researchers have opted to use synthetic datasets due to the scarcity of real-world data. These synthetically generated datasets are large in size and low in labor effort. However, four papers [S6, S11, S16-17] argue that these synthesized datasets can threaten the validity of the proposed methods. It is underscored that the generated data could be significantly different from real-world code smells~\citep{yin2021local}. The generated smelly samples are essentially different from real-world ones that are often more challenging to identify~\citep{liu2019deep}.

\subsection{Limited Generalization Ability}
The limited generalization ability of CSD models is related to the single programming language of the dataset, the limited number of code smell types, and the choice of data source.
Several papers [S1, S3, S7, S12, S15, S22, S24-25, S28, S30, S32, S36] find that the programming language of training data affects model effectiveness. \citet{ramos2022transfer} also find that Python models have lower performance in transfer learning. This performance degradation becomes even more pronounced when evaluated on datasets containing \textit{switch} statements.
In addition, several papers [S1, S5-6, S9, S12, S15-18, S22, S25, S30, S32, S35] have focused on the different code smell types in the datasets. 
For example, three papers [S12, S22, S35] recognize performance degradation when applying their approach to other code smells.
Six papers [S2, S7, S15, S18, S28, S35] indicate that the choice of dataset sources can also limit the generalization ability of models. [S2] points out that using only datasets collected from open-source projects cannot generalize to close-source industrial projects.
The narrowed scope of datasets and classification scenarios may lead to model performance degradation in unseen contexts.

\subsection{Limited Data Accessibility}
The limited data accessibility refers to the unreproducible or unavailable datasets used in CSD.
Many papers [S3, S8, S13-15, S22, S24, S29-30] do not provide access to their source code or the constructed datasets. Some even do not reveal the dataset construction details. For example, [S6] uses private libraries to build datasets. While other researchers cannot access the same datasets to validate or extend the study.
Lack of reproducibility in scientific research means reduced impact of the results~\citep{lewowski2022far}. For example, it is difficult for the industry to trust, invest in, or apply ideas or findings that cannot be replicated in practice. We suggest that future studies should select publicly available and representative data sources when constructing the datasets to ensure that the study is replicable, scalable, and widely applicable.

\subsection{Heavy Expert Dependency}
Heavy Expert Dependency refers to the manual labeling of the code smell datasets described in the previous subsection, which imposes a substantial demand for expertise on the experts~\citep{ho2023fusion}. Many papers [S3, S6, S14, S17-18, S26, S34, S36] have mentioned that data experts should deeply understand the distinct characteristics and intricate concepts underpinning various code smell types. For example, [S17] proposes to train inspectors and enhance their conceptual and cognitive grasp of the code smell domain, evaluating their aptitude to select excellent graduate students for the manual evaluation. Moreover, it is worth noting that the identification of certain complex code smells can pose formidable challenges to researchers. These intricacies can make the task exceedingly difficult. Even experts can find it hard to agree on the presence of a smell sample~\citep{palomba2015landfill}. For example, \citet{bavota2013methodbook} invite 105 experts to evaluate all refactoring suggestions generated by their models and identify whether they agree with the smelly samples. Results show that 44\% of the experts disagree with each other. 

\subsection{Difficulty of Data Labeling}
Data labeling poses unique challenges for code smell detection, including time costs for manual labeling, accuracy issues with automatic approaches, and potential label noise.
Five papers [S5-6, S18, S24, S36] emphasize difficulties with manual and automatic labeling. Manually building labeled datasets is time-consuming due to the extensive effort required [S5]. Though automatic labeling could help with scale, tools have limitations related to predefined heuristics and lower accuracy than human judgments [S6]. Label noise originating during dataset construction also impacts quality. Manual labeling relies on subjective developer perspectives and inconsistent interpretations of smell definitions, which can lead to ambiguous or conflicting labels for the same code [S7, S33, S35]. To experiment on manually validated datasets, [S5] observes significant performance decreases compared to generated data. This highlights the risk of models overfitting to potentially noisy human-generated labels. In summary, both manual and automatic approaches present difficulties that hinder effective labeling at scale. The subjective nature of code smells also makes datasets susceptible to label noise. These challenges point to the need for labeling methods that balance accuracy, efficiency, and consistency in dataset construction.

\subsection{Data Imbalance}
Data imbalance refers to an uneven distribution of code smell samples within datasets, where smelly samples are generally outnumbered by non-smelly code. Such imbalance poses challenges for model training and performance ~\cite{gong2019novel,yu2017learning,feng2021coste}.
For example, the widely-used dataset Qualitas [S2, S10, S21, S25, S27] exhibits a significant imbalance, containing only 33\% smelly samples. Some datasets are even more skewed, with four studies [S5, S11, S16-17] utilizing data comprising just 2\% smelly samples. [S28] indicates severe performance degradation caused by data imbalance. And [S6] highlights imbalance slows down model convergence during training due to the overabundance of non-smelly samples, prolonging the training time. Moreover, models trained on such imbalanced data tend to over-predict samples as non-smelly, which hinders the detection of the underrepresented yet important smelly samples. 

\subsection{Data Redundancy}
Data redundancy refers to duplicate or overlapping samples and features within datasets used for model training ~\cite{li2023impact,jian2019hybrid}.
Sample redundancy means multiple identical code samples exist in the dataset. Several papers [S2, S14-15] observe this occurrence. Redundant samples unnecessarily consume storage and computational resources during training, as they provide no additional value~\citep{xu2021multi, sharma2021code}.
Redundant features refer to excessive or overlapping code smell features within the dataset. Overlapping features can complicate models and hurt performance, as highlighted by [S2]. Such redundancy can arise when feature extraction tools derive similar code smell features from source code. We identify four papers [S2, S13, S21, S27] that initially use such tools to characterize code smells and construct deep learning models based on these extracted features.

\subsection{The Interplay of Challenges}

In previous discussion, we have addressed each identified challenge in isolation to explain their specific impacts and solutions. However, it is crucial to recognize that these challenges often do not occur independently but interplay in ways that can compound their effects. For example, data scarcity often leads to a disproportionately smaller number of smelly instances than non-smelly instances within datasets, thereby exacerbating data imbalance (referenced in [S14, S16]). Similarly, heavy reliance on expert knowledge not only hinders the efficiency of data labeling but also contributes to overall data scarcity, further complicating the challenges (as noted in [S6, S18, S36]). Furthermore, issues such as limited data accessibility combined with high data redundancy can severely limit the generalization ability of the solutions (discussed in [S3, S6, S15, S22, S24, S28, S30]).

To effectively tackle these multifaceted problems, it is apparent that solutions need to be designed with an understanding of these dynamics. Therefore, we propose that a more systematic approach, possibly incorporating integrated solutions, is essential to address the interrelated challenges comprehensively. This holistic perspective is crucial for developing robust and effective strategies to address the complexities.

\section{RQ3 - Solutions Presented in the Literature} \label{RQ3}

\begin{figure*}
    \centering
    \includegraphics[width=0.75\linewidth]{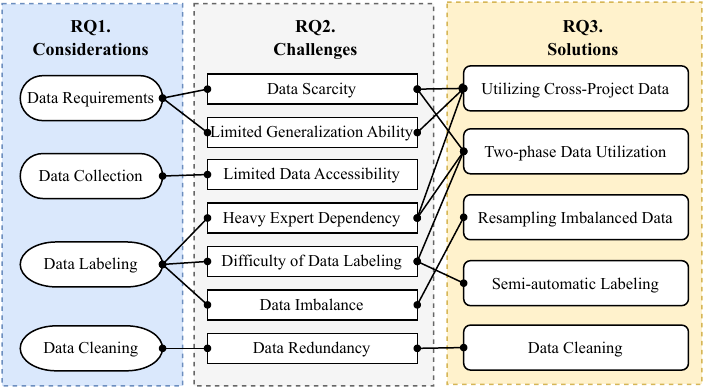}
    \caption{The mappings between three research questions.}
    \label{RQ summary}
\end{figure*}

The previous sections have outlined several key challenges in constructing datasets for DL-based CSD. This section summarizes solutions presented in the literature for mitigating these issues.
Figure~\ref{RQ summary} summarises the solutions we have identified and marks the challenges in RQ2 that they address.
The following analysis examines these proposed approaches, assessing their potential, as well as their limitations.

\subsection{Utilizing Cross-project Data}

\noindent \textit{(Target Challenges in RQ2: Data Scarcity, Limited Generalization Ability, and Heavy Expert Dependency)}

Leveraging data from multiple software projects is an approach for constructing more high-quality datasets ~\cite{yu2018cross,xu2019cross,gong2019novel}. We identify 13 papers [S5-6, S8, S10-11, S14, S16, S24, S28, S31, S34-36] that apply this approach to address challenges related to manual dataset creation efforts. Key benefits of cross-project datasets include improved efficiency, diversity, and generalizability.

Specifically, drawing from various codebases significantly reduces time and resources spent on manual labeling compared to single-project datasets~\citep{barbez2019deep}. This approach also seamlessly accommodates differing languages and smell types to address dataset homogeneity issues~\citep{sharma2021code}. Moreover, validating models across projects further enhances their assessed generalizability. Rather than overfitting to narrow contexts, cross-project datasets support identifying smells in new codebases. This confirms a model's versatility versus the limitations of project-specific evaluations. 

In summary, utilizing multi-project sources presents an effective strategy for constructing more high-quality, diverse, and representative datasets. The enhanced scale, heterogeneity, and generalizability provided by cross-project data help advance research by enabling the development of detection models applicable to broad codebase populations. This solution directly addresses key challenges around dataset construction efforts.

\subsection{Two-phase Data Utilization}
\noindent \textit{(Target Challenges in RQ2: Data Scarcity, Heavy Expert Dependency, and Difficulty of Data Labeling)}

Aiming to enhance model performance, there are ten papers [S2-3, S14, S18, S25, S27, S32, S33-35] employing a two-phase pre-training and fine-tuning approach.

During pre-training, models learn patterns from synthetically generated data. This provides a foundation of code smell characteristics despite potential differences from real data distributions. The primary goal is exposure rather than perfect replication. Subsequently, fine-tuning involves further refining the model using real-world datasets. Adjusting the learnable parameters of deep learning models helps specialize them to real-world code smells.

Research has shown deep learning can cope with noise in training data~\citep{liu2019deep}. Thus, synthetic data facilitates dataset expansion while fine-tuning addresses challenges of low reliability and scarce real-world examples.
Pre-training establishes a general understanding before fine-tuning customizes performance for real-world accuracy~\citep{ren2021exploiting}. Overall, this staged process maximizes the value of generated data. The two-phase approach cultivates models capable of detecting smells across diverse codebases addressing practical scenarios. By integrating synthetic and real data synergistically, this strategy helps advance the field.

\subsection{Resampling Imbalanced Data}
\noindent \textit{(Target Challenges in RQ2: Data Imbalance)}

Data resampling aims to address the imbalance by rebalancing class distributions. The appropriate technique depends on the imbalance severity and requirements.

We identify 14 papers [S4, S12-16, S21-23, S25, S27, S31, S33, S36] exploring resampling. The approach used by most studies [S4, S13, S21, S25, S27, S33] is \underline{S}ynthetic \underline{M}inority \underline{O}ver-sampling \underline{T}echniqu\underline{E} (SMOTE)~\citep{chawla2002smote}. This oversampling technique generates synthetic smelly samples to help balance the datasets. Apart from SMOTE, one study [S23] uses fuzzy sampling, and [S22] develops an automatic refactoring tool to transform non-smelly samples into smelly ones. Five papers [S12, S14-16, S36] apply undersampling to reduce the non-smelly samples.

\subsection{Semi-automatic Labeling}
\noindent \textit{(Target Challenges in RQ2: Difficulty of Data Labeling)}

To address labeling challenges, 14 papers [S2-3, S14, S16, S18, S25, S27-28, S31-36] apply a semi-automatic methodology combining automatic tools with manual validation for high-quality labeled data.

Generally, this approach uses automatic tools to label all samples and verifies the results that are labeled by experts. Specifically, multiple experts chose a subset of samples for individual analysis regarding code smells. The experts perform individual analyses without discussing them with others. Then, Cohen's Kappa is calculated to measure inter-expert agreement. Disagreements are discussed to reach a consensus on a final labeled set. Finally, the manually labeled results are compared to the automatic tools' labeled results. If the differences are negligible, the dataset labeled by the automatic tools for all samples is ultimately adopted.

This approach not only facilitates large-scale labeling but also ensures quality through expert verification. For instance, study S[2] achieves balanced datasets by integrating tool-based advice with expert validation, demonstrating improved robustness in model performance. Similarly, S[16] and S[36] underscore the method's efficacy in refining datasets for enhanced model learning outcomes. Such empirical evidence highlights the semi-automatic approach as both effective and efficient.

\subsection{Data Cleaning}
\noindent \textit{(Target Challenges in RQ2: Data Redundancy)}

The goal of cleaning is to derive datasets optimized for accurate detection. Data cleaning involves removing redundancy, inconsistencies, and irrelevant content. 
Our survey identifies several common cleaning methods, including:
\begin{itemize}
    \setlength{\itemsep}{0pt}
    \setlength{\parsep}{0pt}
    \setlength{\parskip}{0pt}
    \item Comment/blank line removal to filter non-code contextual data [S3, S21, S27].
    \item Missing value imputation or sample removal to handle gaps [S21, S27-28].
    \item Outlier detection/replacement to manage abnormal distributions [S21, S27].
    \item Data type conversion to fix incorrect format [S33].
    \item Feature scaling/normalization to standardize attribute ranges [S18, S27-28].
    \item Feature selection techniques to remove redundancy features [S6, S9-10, S13, S18-21, S25].
    \item De-duplication to remove replicate samples [S27].
\end{itemize}

These techniques facilitate downstream smell feature capture by pruning problematic samples and preparing clean, consistent data. Models can then better learn from focused, high-quality inputs. Notably, S[9] and S[13] have shown that careful feature selection is crucial for model accuracy. Further, S[20] and S[25] demonstrate that reducing feature redundancy not only improves model performance but also reduces computational demands. Additionally, S[33] highlights the significant performance enhancements achievable through data standardization, affirming the critical role of these techniques in preparing high-quality datasets for deep learning. Further refinement of these cleaning practices remains an active area of research to develop high-quality CSD datasets.

\section{Recommendation} \label{recommendation}
Figure~\ref{RQ summary} summarizes our findings on key data preparation considerations, challenges, and potential solutions based on the literature review. This section aims to provide recommendations for researchers and practitioners guided by these results.

\paragraph{Develop Datasets Across Languages and Sources.}
Most papers [S1-12, S15-33, S35] focus on a single programming language, limiting external generalizability and cross-language applicability. Only two papers [S14, S34] examined transfer learning across languages. Manual dataset creation also poses expertise and labor challenges. Automated generation relies on subjective tools restricting new smell detection.
To address these, we recommend utilizing cross-project datasets leveraging multiple codebases. This reduces manual effort while enhancing the diversity of languages and smells represented. Researchers should also focus on expanding language support beyond the dominant Java studies to enable transfer learning assessments.
We further suggest applying a two-phase pre-training and fine-tuning strategy. This marries the benefits of plentiful synthetic data for pre-training with refinement on real-world examples, improving generalizability.

\paragraph{Standardize Cross-study Dataset.}  Lack of consistency hinders reproducibility and progress. Researchers should consider standardizing aspects like identifier naming, feature representations, and metadata tracking across publically available datasets. This will facilitate continued research efforts on cross-dataset challenges. Integrating dataset construction into end-to-end pipelines also helps. Many studies optimized individual preparation phases in isolation. Developing consolidated pipelines covering data sourcing, labeling, cleaning, and modeling could promote co-optimization of these interrelated tasks. Automating pipelines would further decrease manual overhead. In addition, we also recommend setting up centralized data repositories. Finding, preparing, and reusing existing datasets requires significant effort. Researchers should consider developing open centralized repositories and standardized metadata to overcome these barriers.

\paragraph{Adopt Semi-automated Labeling Approaches.}
Dataset labeling plays a critical role in the preparation process. While manual labeling guarantees high accuracy, it requires significant time and expertise that risks scalability issues (i.e., expertise requirements, labor intensity). Meanwhile, fully automated labeling raises accuracy concerns, especially for complex smells.
To address these challenges, we recommend increased utilization of semi-automated labeling approaches. Combining automated prior generation with expert validation, these hybrid methods capture the benefits of both worlds. They considerably reduce manual effort compared to pure manual labeling while maintaining relatively high-quality labels superior to fully automated techniques alone. This makes semi-automated labeling particularly suitable for training deep-learning models targeting complicated smell types.
Additionally, datasets constructed from publicly available sources may not precisely represent industrial contexts due to project-specific differences (i.e., external generalizability challenge). We thus suggest practitioners leverage semi-automated strategies to build customized, organization-focused datasets, improving application scenario reflection. Standardizing such hybrid workflows could advance research reproducibility and industrial adoption of detection solutions.

\paragraph{Enhance Data Transparency and Sharing.}
Current studies face data privacy and replicability challenges. Some datasets solely rely on open-source repositories, introducing subjectivity issues. The inability to fully access or reproduce original private databases also limits validation. We recommend researchers clearly document their full data preparation process. This transparency allows others to comprehensively understand and replicate methodologies. Researchers should also utilize open data platforms to publicly share datasets while preserving privacy. Key metadata around sources and quality assurances enhances usability and trust for the research community. Establishing centralized repositories incentivizing data contributions could help amass more comprehensive benchmark resources. Engaging the industry collaboratively in curating realistic problem snapshots likewise benefits the field.
Standardizing metadata schemas and licensing models promotes long-term data maintenance. Proper governance balances privacy, reproducibility, and continued community-driven progress in solving critical problems. Overall, data availability remains paramount for advancing this impactful research domain.

\paragraph{Establish Data Quality Evaluation Standards.}
Dataset quality impacts the credibility and reproducibility of findings. As our review shows, some papers exhibit class imbalance, limited generalizability, noise, and redundancy. We recommend the research community develop a standardized set of data quality criteria. Metrics should assess the key attributes mentioned above and more. For example, balance criteria could specify acceptable sample size ratios between smells/non-smells. Diversity standards may require code drawn from different projects/domains to ensure representativeness. Noise and redundancy checks aim to flag and remove problematic samples. Researchers should systematically apply the proposed criteria when curating and publishing datasets. Studies could also quantitatively benchmark resources pre/post quality improvements to validate enhancement approaches. Establishing clear quality baselines empowers comparative assessment and continued refinement. It promotes replication by formalizing important methodological aspects. Overall, endorsed evaluation practices can help judiciously manage preparation trade-offs and advance the field through high-confidence shared resources.

\paragraph{Potentials of Large Language Models.}

Large Language Models (LLMs) exhibit excellent zero-shot and in-context learning capabilities, can automatically leverage large-scale data, and are applicable across multiple domains~\citep{gutierrez2022thinking}.
In software engineering, LLMs are increasingly recognized for their utility in various applications. For instance, in vulnerability detection, LLMs assist in identifying potential security risks~\citep{akuthota2023vulnerability}, while in code completion, they facilitate the automatic completion of code fragments based on contextual cues~\citep{roziere2023code}.

Specifically, in the realm of code smell detection, LLMs have the potential to identify common code smell patterns by learning from extensive code samples. This capability can aid programmers in detecting and rectifying issues more efficiently. LLMs offer adaptability across different programming languages and project scales compared to existing techniques. Despite the current absence of research directly combining LLMs with CSD, the potential for LLMs to reduce data labeling workloads and tackle data scarcity challenges significantly is compelling. This makes them a promising technology for addressing the needs of high-quality CSD datasets.

\paragraph{Advancing CSD with Proven Data Techniques.}
Various studies within the software engineering field highlight the importance of robust data preparation. \citet{croft2022data} discuss the challenges of data imbalance and labeling noise, common to our own findings in Code Smell Detection (CSD). They advocate for using class rebalancing and specific data cleaning techniques, such as removing blank lines, non-ASCII characters, comments from code, and duplicate code instances. A unique approach they propose, which differs from our current methodology, is the replacement of user-defined variables and function names with generic tags to reduce code noise further. This strategy could potentially enhance the generalizability of our CSD models by minimizing overfitting to specific code styles or developer idiosyncrasies. According to \citet{yang2022survey}, a classification of the dataset based on data types—such as code data and metric data—is essential for maintaining relevance and accuracy in data analysis. For code data, it is necessary to filter out irrelevant elements while preserving valuable source code and to remove duplicated instances that can skew the analysis. For metric data, normalization is crucial when values span different orders of magnitude, to ensure that no single metric disproportionately influences the model. As explored by \citet{shi2022we}, the development of automatic data cleaning tools represents a significant advancement in handling noisy data within software engineering projects. These tools utilize heuristic rules to automatically identify and remove common issues such as empty functions and duplicated code. Incorporating such tools into our data preparation process could streamline our workflows and improve the quality of our datasets, ultimately leading to more reliable CSD detection models. We believe that integrating cross-disciplinary techniques could improve the overall effectiveness and robustness of CSD models.

\section{Threats to validity} \label{threat}
As with any systematic literature review, this study faces potential threats to validity that could influence results and conclusions. We classify threats based on construct, internal, external, and conclusion validity~\citep{zhou2016map}. While not exhaustive, reporting these threats promotes transparency. The study's context is DL-based code smell detection techniques to establish a foundation for methodological discussions over the specified period - general conclusions require corroborating evidence. We aim to contextualize results by openly reporting on these threats and mitigation efforts.

\paragraph{Construct Validity} is about the connection between the research hypothesis and the findings associated with the RQs. Threats about this category are usually related to the RQs, search strategy, and paper selection process. To mitigate this threat, we create a comprehensive paper selection strategy. 
Specifically, first, we use the search strings and their alternative spellings and synonyms to ensure that most of the key papers are retrieved. For example, some papers do not necessarily include the term deep learning in the title, abstract, or keywords.
We may choose to use the name of the techniques (e.g., RNN or CNN) to ensure that the search string is comprehensive.
Second, we create a paper quality assessment strategy to ensure retrieved papers satisfy the RQs.
Finally, we employ the \textit{snowballing}~\citep{wohlin2014guidelines} approach to obtain further any relevant research that may have been missed.

\paragraph{Internal Validity} is related to the consistency of research findings. 
In this survey, it mainly affects the results of the paper quality assessment and data extraction.
To mitigate this threat, we collaborate with multiple authors to reduce subjectivity in the quality assessment and data extraction processes. Specifically, during the quality assessment phase, one author assesses the quality of all papers, and two other authors validate the results. We will talk about any disagreements and resolve them. Also, one author extracts the data during data extraction, and the other authors validate the extracted data for all the papers. The data are compared, and any conflicts are disscussed and resolved.

\paragraph{External Validity} is related to the generalizability of the reported results. This survey focuses on only one area of software engineering - the data preparation process for DL-based CSD. Therefore, the results can not be generalized to other areas. Furthermore, deep learning is a rapidly evolving field with new techniques being introduced every day~\citep{alazba2023deep}. Our survey is limited to December 2023. Results may not apply to ranges outside the timeline.

\paragraph{Conclusion Validity} is related to the likelihood of reproducing the research and obtaining the same results. To mitigate this threat, we describe the entire research process in detail, including the RQs, the search string, the inclusion/exclusion criteria, the quality assessment form, and the data extraction form. In addition, our findings, i.e., considerations, challenges, and solutions, are based on data extracted from the original papers. We ensure the integrity of our survey and the reality of our findings through rigorous paper selection.

\section{Conclusion} \label{conclusion}
Through a systematic analysis of 36 relevant studies on data preparation approaches for deep learning-based code smell detection until December 2023, our survey illuminates key aspects of the data preparation process. We explore considerations across the data requirements, collection, labeling, and cleaning phases, as well as prevalent challenges and proposed solutions from the literature. 
Key challenges identified include data scarcity, limited generalizability, difficulties in labeling, data imbalance, and redundancy issues. The solutions proposed focused on leveraging cross-project data, two-phase data utilization, semi-automated labeling,  resampling, and data cleaning techniques. 
Based on these results, our primary recommendations are for researchers to establish standardized practices around dataset quality assessment, transparency, and centralized resources. We also recommend techniques for practitioners to construct industrial-strength datasets representative of real-world codebases. 
This systematic review provides a foundation for rigorously evaluating CSD data preparation efforts. Adopting its recommendations aims to foster continued optimizations towards better detection capabilities with real-world application potential. 

\paragraph{Limitations and Future Work}
To address the limitations inherent in a systematic literature review, we acknowledge the absence of empirical validation of the recommendations made in this paper. While our study provides a comprehensive synthesis of available literature on data preparation processes for deep learning-based code smell detection, the conclusions drawn remain hypothetical. Recognizing this, we commit to staying updated on the latest developments in DL-based CSD. We plan to conduct practical experiments and detailed case studies for empirical validation in future studies. Such empirical evidence will be crucial to substantiate the effectiveness of the proposed solutions and further advance the field.

\bibliographystyle{cas-model2-names}

\bibliography{main}

\end{document}